\documentclass[review]{elsarticle}

\usepackage{lineno,hyperref}
\modulolinenumbers[1]

\journal{}

\biboptions{authoryear}

\usepackage{amsmath} 
\usepackage{amssymb}
\usepackage{epstopdf}
\usepackage{graphicx}
\usepackage{eurosym}
\usepackage{color}
\newtheorem{remark}{Remark}[section]
\usepackage{subfigure}

\usepackage[latin1]{inputenc}
\usepackage[english]{babel}

\begin{document}

\begin{frontmatter}

\title{Dynamic compensation and homeostasis: \\ a feedback control perspective}


\author[mymainaddress,myfourthaddress]{Michel Fliess}

\author[mysecondaryaddress,myfourthaddress,myfifthaddress]{C\'{e}dric Join}


\address[mymainaddress]{LIX (CNRS, UMR 7161), \'Ecole polytechnique, 91128 Palaiseau, France.  \newline
        {\tt\small Michel.Fliess@polytechnique.edu}}
\address[mysecondaryaddress]{CRAN (CNRS, UMR 7039), Universit\'{e} de Lorraine, BP 239, \\ 54506 Vand{\oe}uvre-l\`{e}s-Nancy, France \\  {\tt\small cedric.join@univ-lorraine.fr}}
\address[myfourthaddress]{AL.I.E.N. (ALg\`{e}bre pour Identification \& Estimation Num\'{e}riques), \\ 7 rue Maurice Barr\`{e}s, 54330 V\'{e}zelise, France. \newline
        {\tt\small \{michel.fliess, cedric.join\}@alien-sas.com}}
\address[myfifthaddress]{Projet Non-A, INRIA Lille -- Nord-Europe, France}

\begin{abstract}
 ``Dynamic compensation'' is a robustness property where a perturbed biological circuit maintains a suitable output [Karin O., Swisa A., Glaser B., Dor Y., Alon U. (2016). Mol. Syst. Biol., 12: 886]. In spite of several attempts, no fully convincing analysis seems now to be on hand. This communication suggests an explanation via ``model-free control'' and the corresponding ``intelligent'' controllers [Fliess M., Join C. (2013). Int. J. Contr., 86, 2228-2252], which are already successfully applied in many concrete situations. As a byproduct this setting provides also a slightly different presentation of homeostasis, or ``exact adaptation,'' where the working conditions are assumed to be ``mild.'' Several convincing, but academic, computer simulations are provided and discussed.
\end{abstract}


\end{frontmatter}


\newpage
\section{Introduction}\label{intro}
In \emph{systems biology}, \textit{i.e.}, an approach of growing importance to theoretical biology (see, \textit{e.g.}, \cite{alon0,klipp,kremling}), basic control notions, like feedback loops, are becoming more and more popular (see, \textit{e.g.}, \cite{murray,cowan,cosen,vecc2}). This communication intends to show that a peculiar feedback loop  permits to clarify the concept of \emph{dynamic compensation} (\emph{DC}) of biological circuits, which was recently introduced by \cite{karin}. DC is a robustness property. It implies, roughly speaking, that biological systems are able of maintaining a suitable output despite environmental fluctuations. As noticed by \cite{karin} such a property arises naturally in physiological systems. The DC of blood glucose, for instance, with respect to variation in insulin sensitivity and insulin secretion is obtained by controlling the functional mass of pancreatic beta cells.  

The already existing and more restricted \emph{homeostasis}, or \emph{exact adaptation}, deals only with constant reference trajectories, \textit{i.e.}, setpoints. It is achievable via an {\em integral} feedback (see, \textit{e.g.}, \cite{alon,briat,miao,stelling,yi})
\begin{remark}\label{rem}
PIDs (see, \textit{e.g.}, \cite{murray,od}) read:
\begin{equation}\label{pid}
u = K_P e + K_I \int e + K_D \dot{e}
\end{equation}
where 
\begin{itemize}
\item $u$, $y$, $y^\star$ are respectively the control and output variables, and the reference trajectory.
\item $e = y - y^\star$ is the tracking error,
\item $K_P, K_I, K_D \in \mathbb{R}$ are the tuning gains.
\end{itemize}
To the best of our knowledge, they are, strangely enough, more or less missing in the literature on theoretical biology,\footnote{This is of course less the case in \emph{synthetic biology}, \textit{i.e.}, a blending between biology and engineering (see, \textit{e.g.}, \cite{vecc1} and the references therein).} although they lead to the most widely used control strategies in industry.

From $K_P = K_D = 0$ in Equation \eqref{pid}, the following integral feedback is deduced:
\begin{equation}\label{i}
u = K_I \int e
\end{equation}
Compare Equation \eqref{i} with the references above on homeostasis, and \cite{somvanshi}. See \cite{alinea}, and the references therein, for an application to ramp metering on freeways in order to avoid traffic congestion.
\end{remark}
Conditions for DC have already been investigated by several authors: \cite{karin1,karin0,sontag, villa}. Parameter identification, which plays a key rôle in most of those studies, leads, according to the own words of \cite{karin0}, to some kind of ``discrepancy,'' which is perhaps not yet fully cleared up. We suggest therefore another roadmap, \textit{i.e.},  
\emph{intelligent} feedback controllers as defined by \cite{ijc}. Many concrete applications have already been developed all over the world. Some have been patented. The bibliography contains for obvious reasons only recent works in biotechnology: \cite{bara,siaap,toulon,siam,med16}.\footnote{A rather comprehensive bibliography of concrete applications is provided by \cite{bldg}.} 

An unexpected byproduct is derived from Remark \ref{rem} and the comparison in \cite{alinea} between Equation \eqref{i} and our \emph{intelligent proportional} controller (\cite{ijc}). Exact adaptation means now a ``satisfactory'' behavior thanks to the feedback loop defined by Equation \eqref{i} when the working conditions are ``mild.'' The result by \cite{karin} via a mechanism for DC based on known hormonal circuit reactions, which states that exact adaptation does not guarantee dynamical compensation, remains therefore valid in this new context.


This exploratory research report is organized as follows. Intelligent controllers 
are summarized in Section \ref{I}, where the connection between classic PIs and intelligent proportional controllers is also presented. Section \ref{ALIP}, which is hevily influenced by \cite{alinea}, defines dynamic compensation and exact adaptation. Section \ref{numer} displays various convincing, but academic, computer experiments. Some concluding remarks may be found in Section \ref{con}.

\section{Model-free control and intelligent controllers}\label{I}
See \cite{ijc} for full details. 
\subsection{Generalities}\label{loop}
\subsubsection{The ultra-local model}\label{ulm}
The poorly known global description of the plant, which is assumed for simplicity's sake to be SISO (single-input single output),\footnote{For a multivariable extension, see, \textit{e.g.}, \cite{toulon,its}.} is replaced by the \emph{ultra-local model}:
\begin{equation}
\boxed{y^{(\nu)} = F + \alpha u}
\label{1}
\end{equation}
where:
\begin{itemize}
\item the control and output variables are respectively $u$ and $y$;
\item the derivation order $\nu$ is often equal to $1$, sometimes to $2$; in practice $\nu \geq 3$ has never been encountered;
\item the constant $\alpha \in \mathbb{R}$ is chosen by the practitioner such that $\alpha u$ and
$y^{(\nu)}$ are of the same magnitude; therefore $\alpha$ does not need to be precisely estimated.
\end{itemize}
The following comments might be useful:
\begin{itemize}
\item Equation \eqref{1} is only valid during a short time lapse and must be continuously updated;
\item $F$ is estimated via the knowledge of the control and output variables $u$ and $y$;
\item $F$ subsumes not only the system structure, which most of the time will be nonlinear, but also
any external disturbance.
\end{itemize}
\subsubsection{Intelligent controllers}
Set $\nu = 2$. Close the loop with the following \emph{intelligent proportional-integral-derivative controller}, or \emph{iPID},
\begin{equation}\label{ipid}
u = - \frac{F - \dot{y}^\ast + K_P e + K_I \int e + K_D \dot{e}}{\alpha}
\end{equation}
where:
\begin{itemize}
\item $e = y - y^\star$ is the tracking error,
\item $K_P, K_I, K_D \in \mathbb{R}$ are the tuning gains.
\end{itemize}
When $K_I = 0$, we obtain the \emph{intelligent proportional-derivative controller}, or \emph{iPD},
\begin{equation}\label{ipd}
u = - \frac{F - \dot{y}^\ast + K_P e + K_D \dot{e}}{\alpha}
\end{equation}
When $\nu = 1$ and $K_I = K_D =0$, we obtain the \emph{intelligent proportional controller}, or \emph{iP}, which is the most important one,
\begin{equation}\label{ip}
\boxed{u = - \frac{F - \dot{y}^\ast + K_P e}{\alpha}}
\end{equation}
Combining Equations \eqref{1} and \eqref{ip} yields:
\begin{equation}\label{dyna}
\dot{e} + K_P e = 0
\end{equation}
where $F$ does not appear anymore. The tuning of $K_P$ is therefore straightforward. 
\begin{remark}
See \cite{iste} for a comment on those various controllers.
\end{remark}

\subsubsection{Estimation of $F$}\label{F}
Assume that $F$ in Equation \eqref{1} is ``well'' approximated by a piecewise constant function $F_{\text{est}} $.\footnote{This is a weak assumption (see, \textit{e.g.}, \cite{bourbaki}).} The estimation techniques below are borrowed 
from \cite{sira1,sira2} and \cite{sira}. Let us summarize two types of computations:
\begin{enumerate}
\item Rewrite Equation \eqref{1}  in the operational domain (see, \emph{e.g.}, \cite{yosida}): 
$$
sY = \frac{\Phi}{s}+\alpha U +y(0)
$$
where $\Phi$ is a constant. We get rid of the initial condition $y(0)$ by multiplying both sides on the left by $\frac{d}{ds}$:
$$
Y + s\frac{dY}{ds}=-\frac{\Phi}{s^2}+\alpha \frac{dU}{ds}
$$
Noise attenuation is achieved by multiplying both sides on the left by $s^{-2}$, since integration with respect to time is a lowpass filter. It yields in the time domain the realtime estimate, 
thanks to the equivalence between $\frac{d}{ds}$ and the multiplication by $-t$,
{\small
\begin{equation}\label{integral}
{ F_{\text{est}}(t)  =-\frac{6}{\tau^3}\int_{t-\tau}^t \left\lbrack (\tau -2\sigma)y(\sigma)+ \alpha\sigma(\tau -\sigma)u(\sigma) \right\rbrack d\sigma }
\end{equation}
}
where $\tau > 0$ might be quite small. This integral may of course be replaced in practice by a classic digital filter.
\item Close the loop with the iP \eqref{ip}. It yields:
$$
F_{\text{est}}(t) = \frac{1}{\tau}\left[\int_{t - \tau}^{t}\left(\dot{y}^{\star}-\alpha u
- K_P e \right) d\sigma \right] 
$$
\end{enumerate}

\begin{remark}
From a hardware standpoint, a real-time implementation  of our intelligent controllers is also cheap and easy (\cite{nice}).
\end{remark}

\subsection{PI and iP}\label{PiP}
Consider the classic continuous-time PI controller
\begin{equation}\label{cpi}
  u (t) = k_p e(t) + k_i \int e(\tau) d\tau
\end{equation}
A crude sampling of the integral $\int e(\tau) d\tau$ through a
Riemann sum ${\cal{I}}(t)$ leads to
$$
\int e(\tau) d\tau \simeq  {\cal{I}}(t) = {\cal{I}}(t-h) + h e(t)
$$
where $h$ is the sampling interval. The corresponding discrete form
of Equation \eqref{cpi} reads:
$$
u(t) = k_p e(t) + k_i {\cal{I}}(t) = k_p e(t) + k_i {\cal{I}}(t-h) + k_i h e(t)
$$
Combining the above equation with $$u(t-h) = k_p e(t-h) + k_i
{\cal{I}}(t-h)$$ yields
\begin{equation}
\label{eqPIRiemannDiscrSix} u(t) = u(t - h) + k_p \left( e(t) - e(t
- h) \right) + k_i h e(t)
\end{equation}

\begin{remark}
A trivial sampling of the ``velocity form'' of Equation \eqref{cpi}
\begin{equation}\label{cpid}
\dot{u} (t) = k_p \dot{e}(t) + k_i e(t)
\end{equation}
yields
$$
\dfrac{u(t) - u(t - h)}{h} =  k_p  \left(\dfrac{e(t) - e(t -
h)}{h}\right) + k_i  e(t)
$$
which is equivalent to Equation \eqref{eqPIRiemannDiscrSix}.
\end{remark}

Replace in Equation \eqref{ip} $F$ by ${\dot y}(t) - \alpha u (t-h)$ and therefore by
\begin{equation*}
\frac{y(t) - y(t-h)}{h} - \alpha u (t-h)
\label{FestBrutal}
\end{equation*}
It yields
\begin{equation}
\label{eqDiscr_i-POne} u (t) = u (t - h) - \frac{e(t) -
e(t-h)}{h\alpha} - \dfrac{K_P}{\alpha}\, e(t)
\end{equation}

{\bf FACT}.- Equations \eqref{eqPIRiemannDiscrSix} and
\eqref{eqDiscr_i-POne} become {\bf identical} if we set
\begin{align}
\label{eqPI_i-P_corresp} k_p &= - \dfrac{1}{\alpha h}, \quad k_i =
-\dfrac{K_P}{\alpha h}
\end{align}
\begin{remark}
This path breaking result was first stated by \cite{equiv}:
\begin{itemize}
\item It is straightforward to extend it to the same type of equivalence between PIDs and iPDs.
\item It explains apparently for the first time the ubiquity of PIs and PIDs in the industrial world. 
\end{itemize}
\end{remark}

\section{Exact adaptation and dynamic compensation}\label{ALIP}
Equation \eqref{cpid} shows that integral and proportional-integral controllers are close when 
\begin{enumerate}
\item $\dot{e}$ remains small,
\item the reference trajectory $y^\ast$ starts at the initial condition $y(0)$ or, at least, at a point in a neighborhood,
\item the  measurement noise corruption is low.
\end{enumerate} 
The following conditions might be helpful:
\begin{itemize}
\item the reference trajectory $y^\ast$ is ``slowly'' varying, and starts at the initial condition $y(0)$ or, at least, at a point in its neighborhood,\footnote{Setpoints are therefore recovered.}
\item the disturbances and the corrupting noises are rather mild.
\end{itemize}
Then the performances of the integral controller should be decent: this is \emph{exact adaptation}, or \emph{homeostasis}. When the above conditions are not satisfied, \emph{dynamic compensation} means that one at least of the feedback loops in Section \ref{loop} is \emph{negative}, \textit{i.e.}, fluctuations around the reference trajectory due to perturbations and input changes are attenuated.\footnote{The wording ``negative feedback'' is today common in biology, but, to some extent, neglected in engineering, where it was quite popular long time ago (see, \textit{e.g.}, \cite{kupf}). Historical details are given by \cite{zeron} and \cite{vecc2}.}

\section{Two computer experiments} \label{numer}
The two academic examples below provide easily implementable numerical examples. They are characterized by the following features:
\begin{itemize}
\item $K_I = 0.5$ (resp. $K_I = 1$) for the integral feedback in the linear (resp. nonlinear) case.
\item  $\alpha = K_P = 1$ for the the iP \eqref{ip} in both cases.
\item The sampling period is $10$ms. 
\item In order to be more realistic, the output is corrupted additively by a zero-mean white Gaussian noise of standard deviation $0.01$.
\end{itemize}
\subsection{Linear case}
\begin{figure}
\vspace{.20cm}\center
\subfigure[Control]{\rotatebox{-0}{\includegraphics[width=.49375\columnwidth]{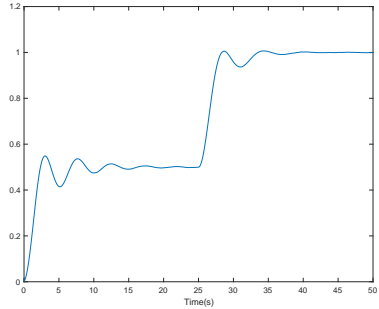}}}
\subfigure[Output]{\rotatebox{-0}{\includegraphics[width=.49375\columnwidth]{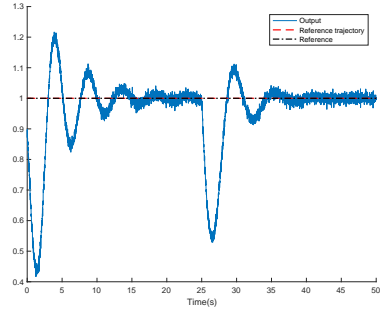}}}
\vspace{-.0cm} \caption{Integral feedback, constant reference trajectory, control efficiency loss \label{IC}}
\end{figure}

\begin{figure}
\vspace{.20cm}\center
\subfigure[Control]{\rotatebox{-0}{\includegraphics[width=.49375\columnwidth]{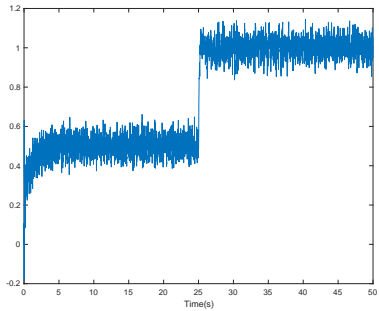}}}
\subfigure[Output]{\rotatebox{-0}{\includegraphics[width=.49375\columnwidth]{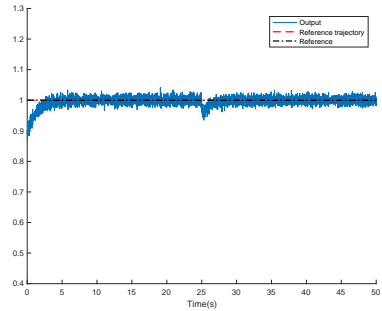}}}
\vspace{-.0cm} \caption{iP, constant reference trajectory, control efficiency loss \label{CSMC}}
\end{figure}

\begin{figure}
\vspace{.20cm}\center
\subfigure[Control]{\rotatebox{-0}{\includegraphics[width=.49375\columnwidth]{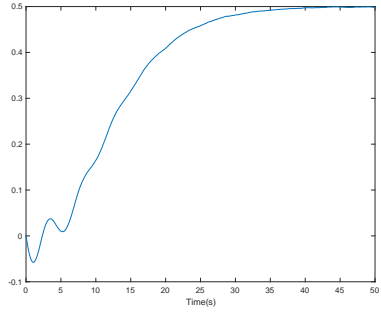}}}
\subfigure[Output]{\rotatebox{-0}{\includegraphics[width=.49375\columnwidth]{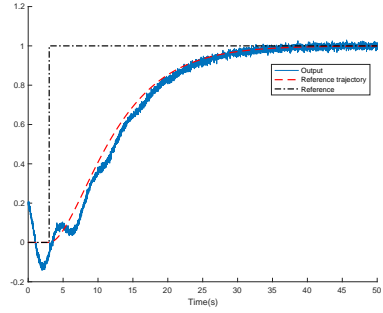}}}
\vspace{-.0cm} \caption{Integral feedback, slow connection \label{IL}}
\end{figure}

\begin{figure}
\vspace{.20cm}\center
\subfigure[Control]{\rotatebox{-0}{\includegraphics[width=.49375\columnwidth]{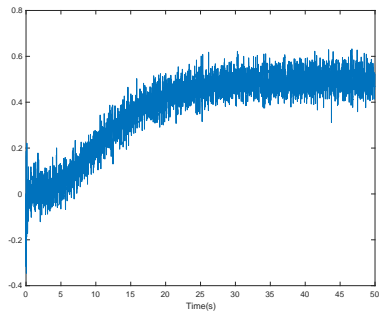}}}
\subfigure[Output]{\rotatebox{-0}{\includegraphics[width=.49375\columnwidth]{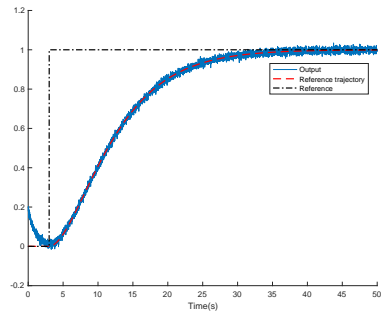}}}
\vspace{-.0cm} \caption{iP, slow connection \label{CSML}}
\end{figure}

\begin{figure}
\vspace{.20cm}\center
\subfigure[Control]{\rotatebox{-0}{\includegraphics[width=.49375\columnwidth]{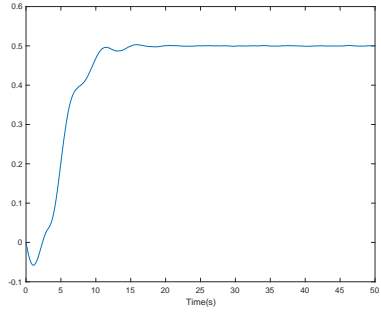}}}
\subfigure[Output]{\rotatebox{-0}{\includegraphics[width=.49375\columnwidth]{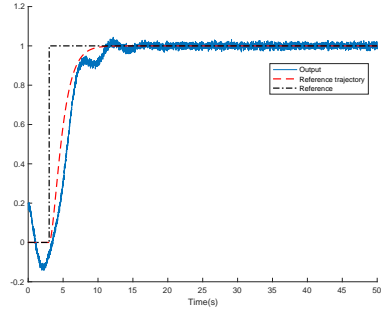}}}
\vspace{-.0cm} \caption{Integral feedback, fast connection \label{IR}}
\end{figure}

\begin{figure}
\vspace{.20cm}\center
\subfigure[Control]{\rotatebox{-0}{\includegraphics[width=.49375\columnwidth]{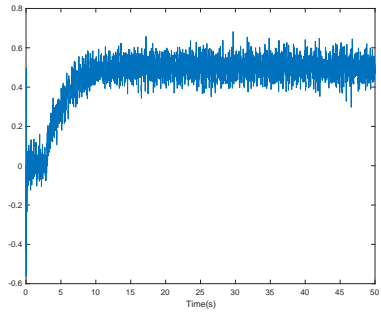}}}
\subfigure[Output]{\rotatebox{-0}{\includegraphics[width=.49375\columnwidth]{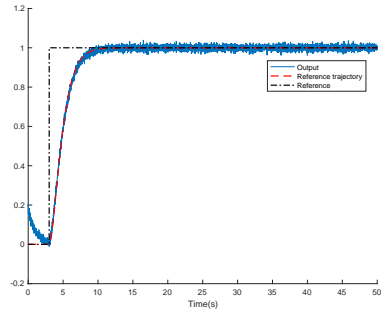}}}
\vspace{-.0cm} \caption{iP, fast connection \label{CSMR}}
\end{figure}

\begin{figure}
\vspace{.20cm}\center
\subfigure[Control]{\rotatebox{-0}{\includegraphics[width=.49375\columnwidth]{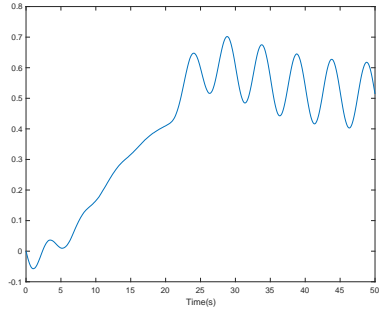}}}
\subfigure[Output]{\rotatebox{-0}{\includegraphics[width=.49375\columnwidth]{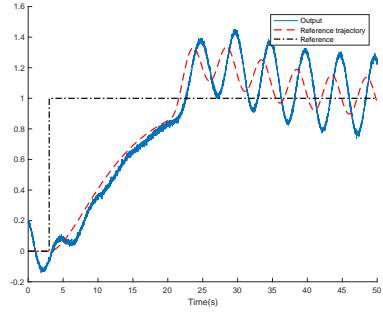}}}
\vspace{-.0cm} \caption{Integral connection, complex reference trajectory \label{IS}}
\end{figure}

\begin{figure}
\vspace{.20cm}\center
\subfigure[Control]{\rotatebox{-0}{\includegraphics[width=.49375\columnwidth]{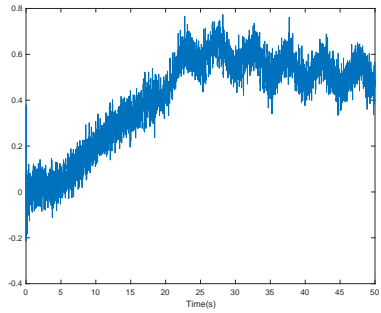}}}
\subfigure[Output]{\rotatebox{-0}{\includegraphics[width=.49375\columnwidth]{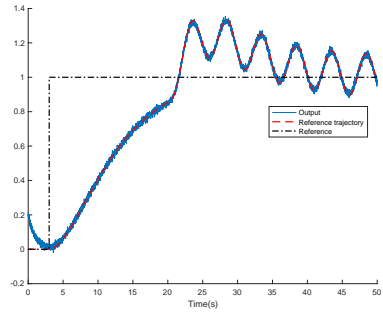}}}
\vspace{-.0cm} \caption{iP, complex reference trajectory  \label{CSMS}}
\end{figure}

Consider the input-output system defined by the transfer function
$$
\frac{2(s+1)}{s^2+s+1}
$$
Several reference trajectories are examined:
\begin{enumerate}
\item[(i)] Setpoint and $50$\% efficiency loss of the actuator: see Figures \ref{IC} see \ref{CSMC}.
 \item[(ii)]  Slow connection between two setpoints:  see Figures \ref{IL} and \ref{CSML}.
\item[(iii)] Fast connection: see Figures \ref{IR} and \ref{CSMR}.
 \item[(iv)] Complex reference trajectory: see Figures \ref{IS} and \ref{CSMS}.
\end{enumerate}
In the first scenario, the control efficiency loss  is attenuated much faster by the iP than by the integral feedback. The behaviors of the integral feedback and the iP with respect to a slow connection are both good and cannot be really distinguished. The situation change drastically with a fast connection and a complex reference trajectory: the iP becomes vastly superior to the integral feedback. Exact adaptation works well only in the second scenario, whereas dynamic compensation yields always excellent results.


\subsection{Nonlinear case}
\begin{figure}
\vspace{.20cm}\center
\subfigure[Control]{\rotatebox{-0}{\includegraphics[width=.49375\columnwidth]{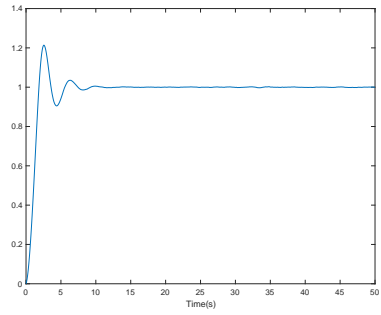}}}
\subfigure[Output]{\rotatebox{-0}{\includegraphics[width=.49375\columnwidth]{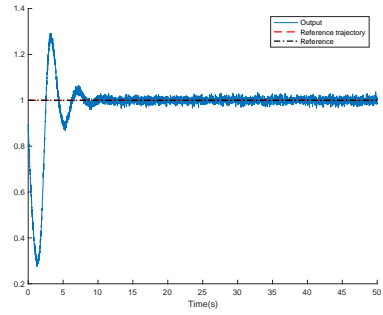}}}
\vspace{-.0cm} \caption{Integral feedback, constant reference trajectory, without any perturbation\label{NIC}}
\end{figure}

\begin{figure}
\vspace{.20cm}\center
\subfigure[Control]{\rotatebox{-0}{\includegraphics[width=.49375\columnwidth]{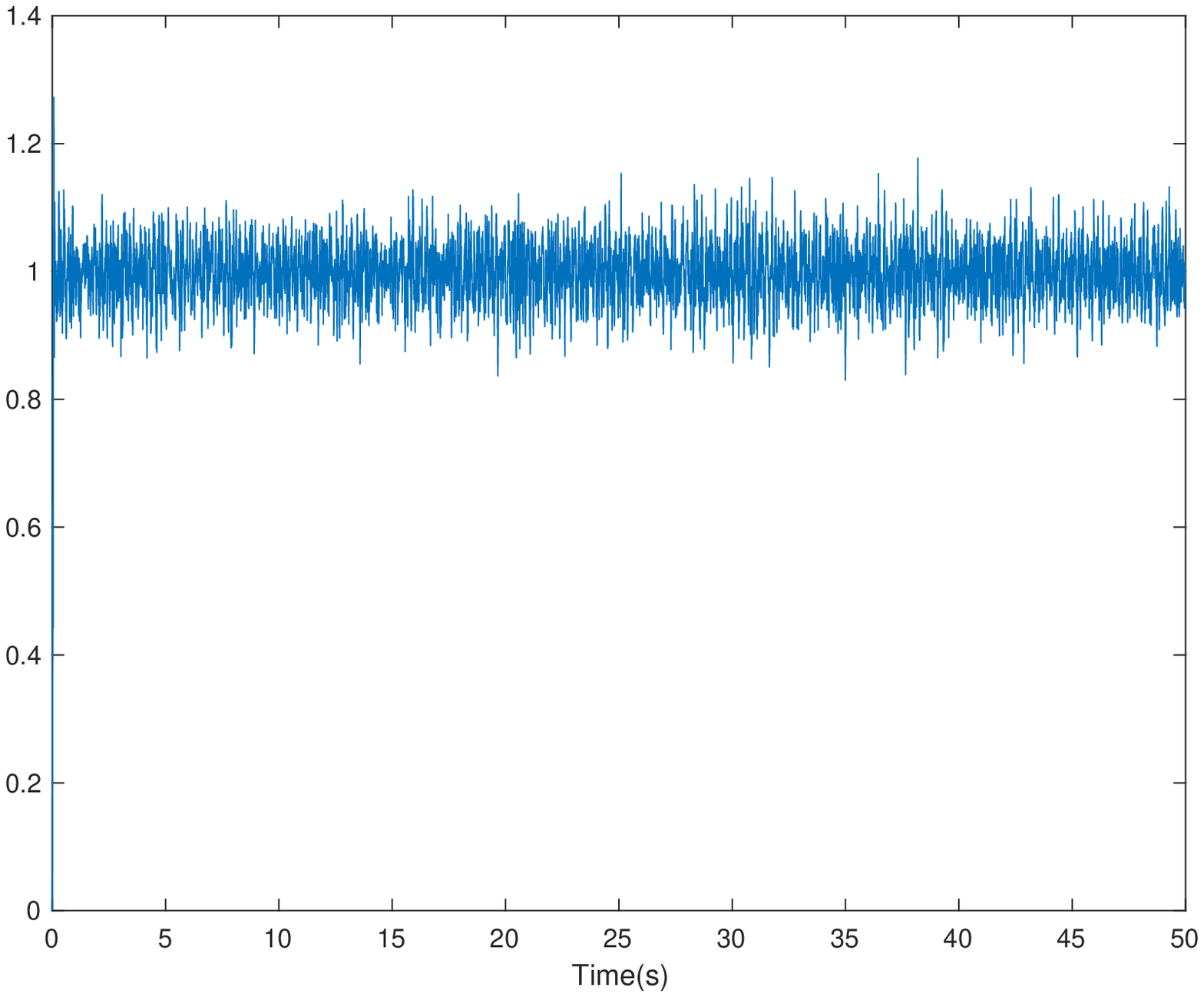}}}
\subfigure[Output]{\rotatebox{-0}{\includegraphics[width=.49375\columnwidth]{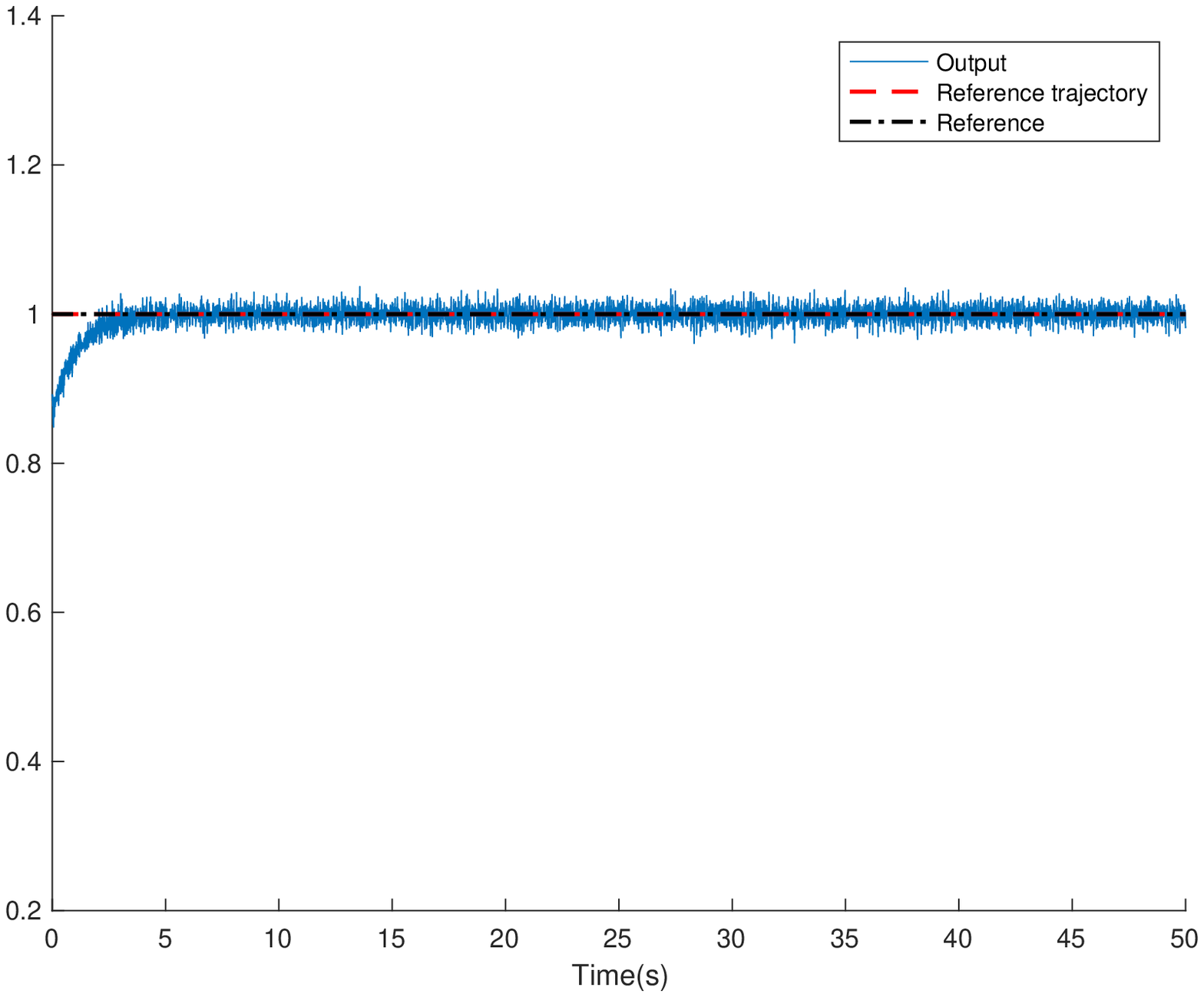}}}
\vspace{-.0cm} \caption{iP, constant reference trajectory, without any perturbation \label{NCSMC}}
\end{figure}

\begin{figure}
\vspace{.20cm}\center
\subfigure[Control]{\rotatebox{-0}{\includegraphics[width=.49375\columnwidth]{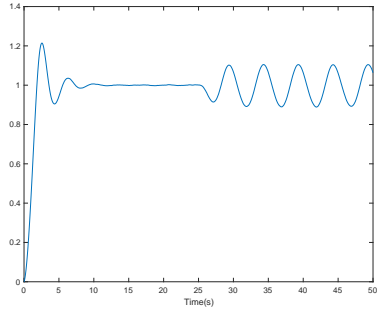}}}
\subfigure[Output]{\rotatebox{-0}{\includegraphics[width=.49375\columnwidth]{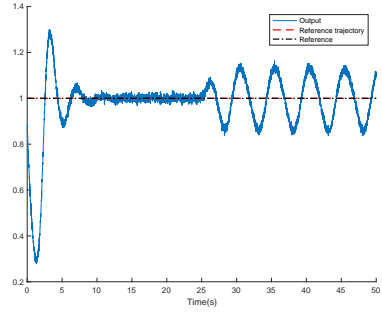}}}
\vspace{-.0cm} \caption{Integral feedback, constant reference trajectory, with perturbation \label{NICP}}
\end{figure}

\begin{figure}
\vspace{.20cm}\center
\subfigure[Control]{\rotatebox{-0}{\includegraphics[width=.49375\columnwidth]{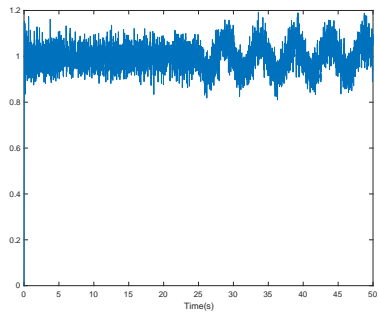}}}
\subfigure[Output]{\rotatebox{-0}{\includegraphics[width=.49375\columnwidth]{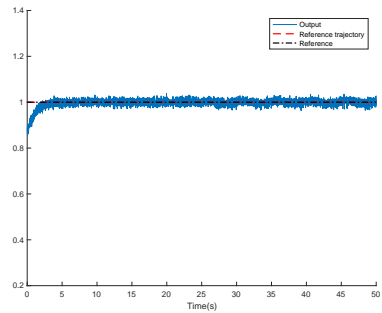}}}
\vspace{-.0cm} \caption{iP, constant reference trajectory, with perturbation \label{NCSMCP}}
\end{figure}

\begin{figure}
\vspace{.20cm}\center
\subfigure[Control]{\rotatebox{-0}{\includegraphics[width=.49375\columnwidth]{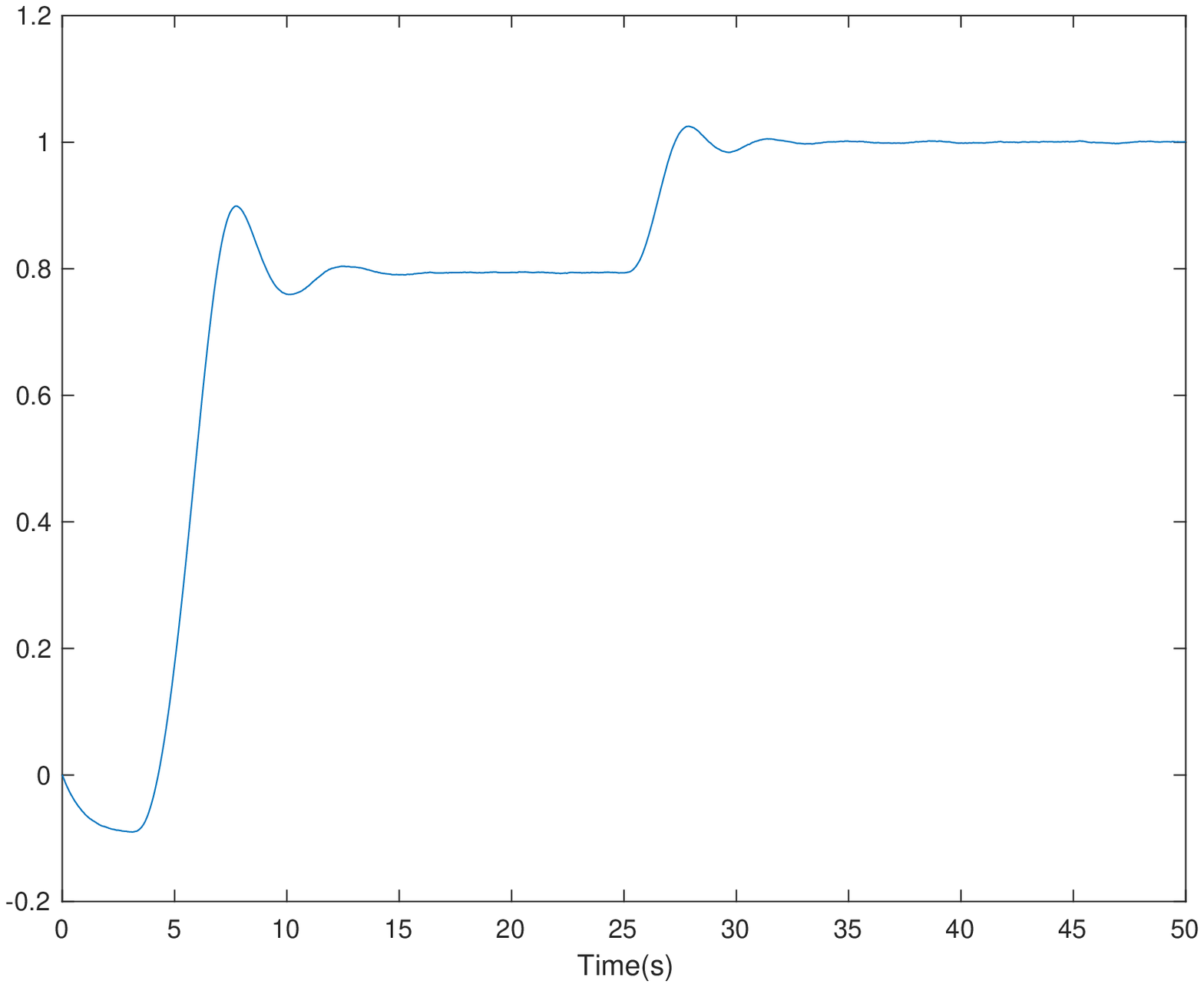}}}
\subfigure[Output]{\rotatebox{-0}{\includegraphics[width=.49375\columnwidth]{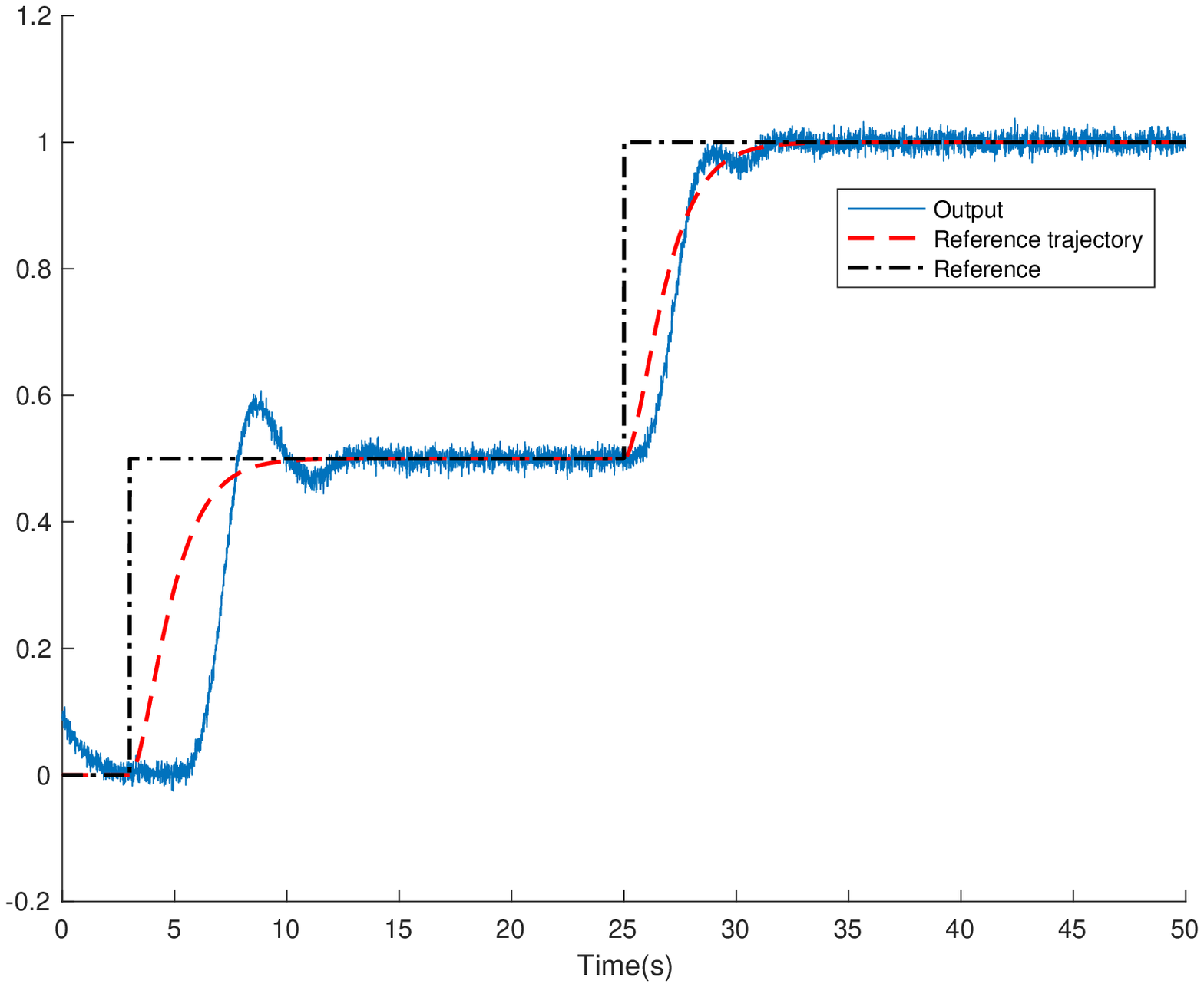}}}
\vspace{-.0cm} \caption{Integral feedback, non-constant reference trajectory, without any perturbation \label{NIR}}
\end{figure}

\begin{figure}
\vspace{.20cm}\center
\subfigure[Control]{\rotatebox{-0}{\includegraphics[width=.49375\columnwidth]{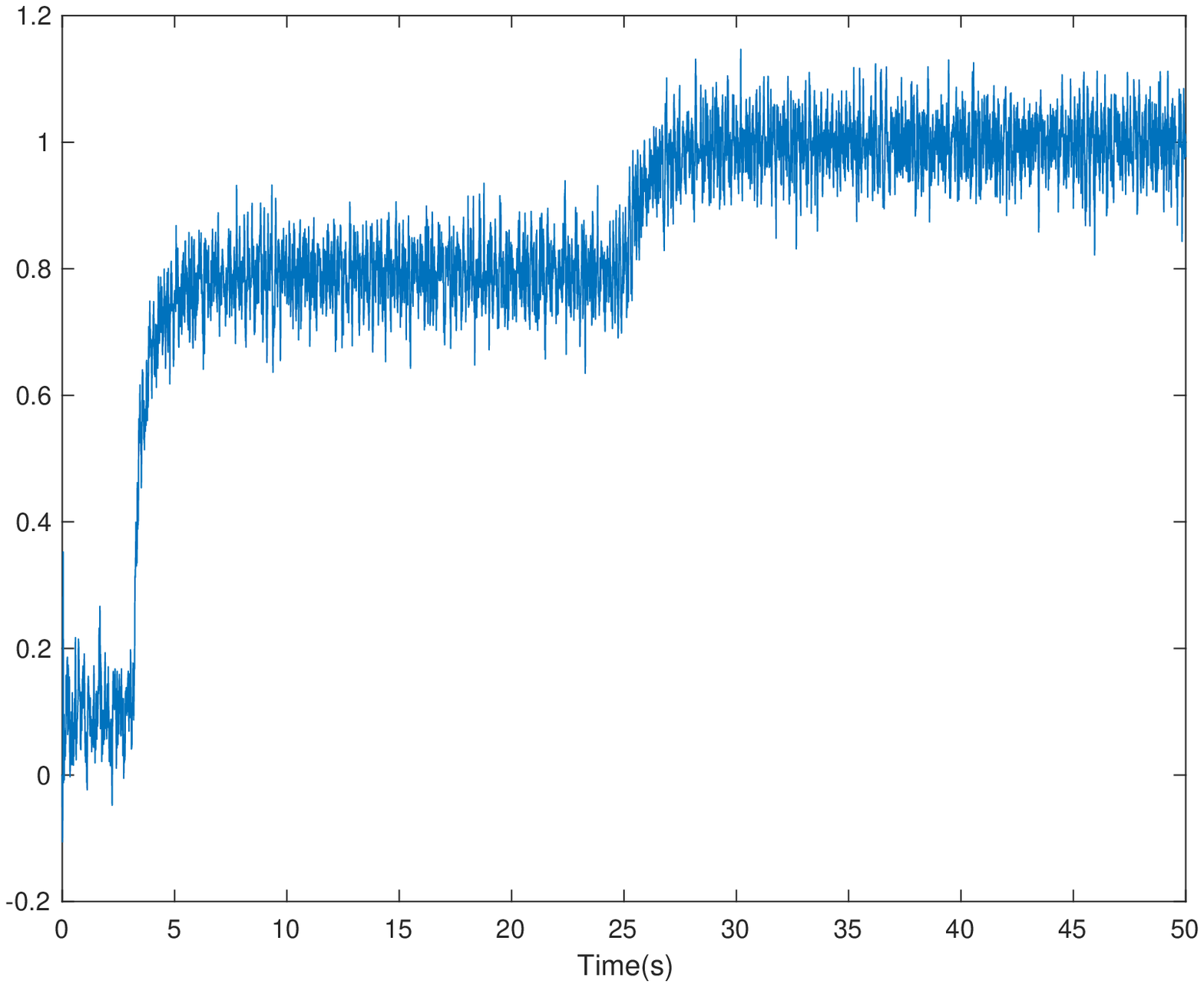}}}
\subfigure[Output]{\rotatebox{-0}{\includegraphics[width=.49375\columnwidth]{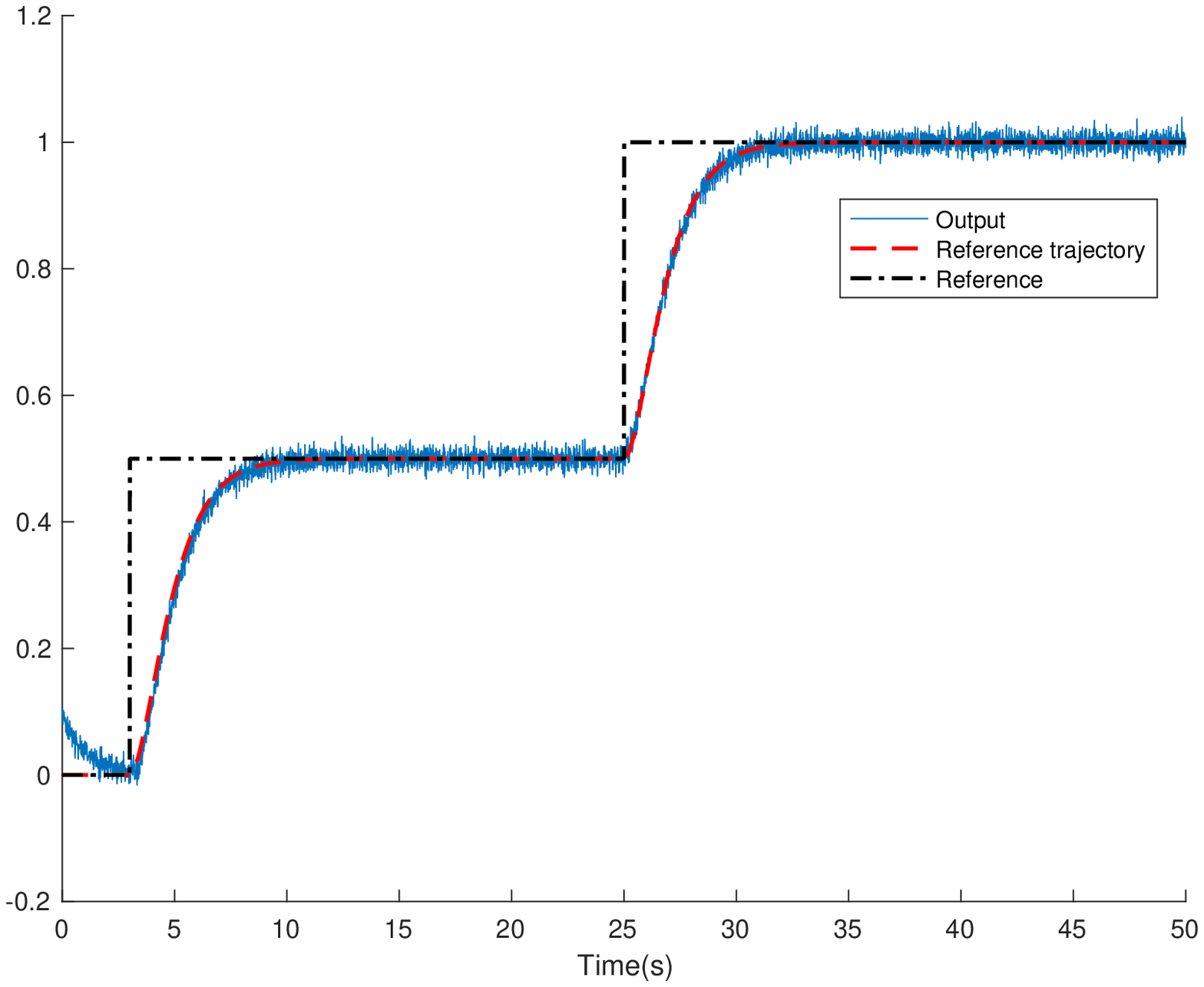}}}
\vspace{-.0cm} \caption{iP, non-constant reference trajectory, without any perturbation \label{NCSMR}}
\end{figure}

Consider
$$
\dot{y} + y=u^3 + P_{\text{pert}}
$$
where $P_{\text{pert}}$ is a perturbation. Introduce three scenarios: 
\begin{itemize}
\item[(i)] Setpoint without any perturbation, \textit{i.e.}, $P_{\text{pert}}=0$: see Figures \ref{NIC} and \ref{NCSMC}.
\item[(ii)] Setpoint with a sine wave perturbation which starts at $t=25$s, \textit{i.e.}, $P_{\text{pert}} (t) = 0.2\sin (\frac{2\pi}{5}(t-25))$ if $t\geq 25$s, and $P_{\text{pert}} (t) = 0$ if $t \leq 25$s: see Figures \ref{NICP} and \ref{NCSMCP}. 
  \item[(iii)] Non-constant reference trajectory without any perturbation, \textit{i.e.}, $P_{\text{pert}}=0$: see Figures \ref{NIR} and \ref{NCSMR}.
\end{itemize}
A clear-cut superiority  of the iP with respect to the integral feedback is indisputable. The behavior of dynamic compensation (resp. exact adaptation) is always (resp. never) satisfactory.
\begin{remark}
Do not believe that integral feedbacks are never adequate if nonlinearities occur. See
\begin{itemize}
\item an example related to ramp metering in \cite{alinea}, 
\item theoretical investigations in \cite{sontag0}.
 \end{itemize}
\end{remark}




\section{Conclusion}\label{con}
 In order to be fully convincing, this preliminary annoucement on homeostasis extensions needs of course to exhibit true biological examples. 
 In our context noise corruption is also a hot topic (see, \textit{e.g.}, \cite{briat,sun}). The estimation and identification techniques sketched in Section \ref{loop} might lead to a better understanding (see also \cite{cras,arima}).



\end{document}